\\

Transition to the Most Probable Kinetic State

in a Pre-Steady State System

Brian K. Davis

(Research Foundation of Southern California Inc.

La Jolla CA 92037, U.S.A.)

e-mail: briand@fisher.biz.usyd.edu.au

Short title: Kinetic time asymmetry

Abbreviation: DNP, 2,4-dinitrophenol; DNPP, 2,4-dinitrophenyl phosphate

1 Figure

\\

A system containing a pre-steady state standard (non-autocatalytic) reaction, with multiple paths, evolves toward a kinetic state with the minimum attainable activation free energy. Displacement of the path frequency distribution in this transition was shown to minimise the affinity linked to this change in activation free energy. In damping this scalar force, a standard system is driven along a path of least action, as previously established for a system of competing autocatalytic reactions. A kinetic source of time asymmetry arises within the system, as the activation affinity moves the system toward the most probable distribution of reaction paths. As the functions of state are not changed by path displacement, a change of kinetic state cannot produce chemical work. This generalises the notion of force to a scalar quantity responsible for a displacement that does not yield work or heat. Spectrophotometric observations reported on the transition to steady state kinetics during dinitrophenyl phosphate phosphorolysis confirmed that time variations in the activation affinity are non-positive.

\\

1. **Introduction**

A deviation from the mean activation free energy for nucleotide condensation in a heterogeneous population of competitively replicating polymers results in a path-centred force (Davis, 1994, 1996) that shifts the polymer frequency distribution and increases the rate of synthesis. Elimination of polymer species with suboptimal template competence accounts for this increase (Kramer et al., 1974; Eigen et al., 1989). In effect, nucleotide condensation tends to take place by progressively more rapid reaction paths, or, equivalently, by more probable transition states. The system thus evolves with time in the direction of the most probable kinetic state.

This leads to the proposition that two independent sets of forces can orient a chemical system in time. A scalar force that increases internal entropy accounts for the thermodynamic orientation of a process in time. The time asymmetry in competitive replication, and putatively in other pre-steady state chemical systems (Laidler, 1987), however, arises from the spontaneous displacement of path frequencies between a reactant and product state, which remain unaltered. In addition to thermodynamic forces responsible for changes of state, therefore, a chemical system also apparently contains path-centred forces that move it toward the most probable set of paths between states.

It appeared of interest to establish that the proposed theory can be extended to a change of kinetic state by a standard reaction. This would reveal whether changes in a standard pre-steady state system (i) damp the activation affinity at a maximal rate, and (ii) yield a path of least action, as in a set of autocatalytic reactions. In contrast to evolution among competing autocatalytic reactions, where, in principle, only the fastest reaction path is ultimately retained (Davis, 1996), all available paths are retained during the transition to steady state kinetics by a standard reaction. Moreover, steady state kinetics are achieved rapidly in the latter. A transition interval of less than 200 ms, for instance, occurred in an enzyme catalysed reaction (Trentham & Gutfreund, 1968) considered here. This compares with a transition interval of nearly 3 hr in the RNA evolution experiment (Kramer et al., 1974) investigated previously. For the same principles to govern a change of kinetic state in two systems with such significant qualitative and quantitative differences, they would have to apply in general.

2. **Transition to Steady State and the Activation Affinity**

A standard reaction with more than one transition state, T(i), forms a set of parallel, consecutive reactions,

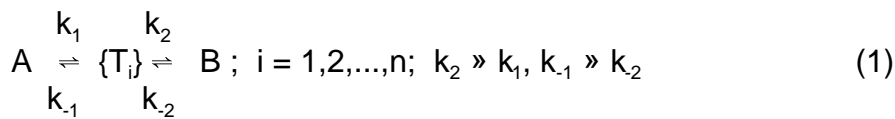

$$A \underset{k_{-1}}{\overset{k_1}{\rightleftharpoons}} \{T_i\} \underset{k_{-2}}{\overset{k_2}{\rightleftharpoons}} B \; ; \; i = 1,2,...,n; \; k_2 \gg k_1, k_{-1} \gg k_2 \qquad\qquad (1)$$

Each reaction path is seen to produce and consume an unstable intermediate, with rate coefficients $k_i$. The frequency distribution of the set $\{T(i)\}$ determines the 'kinetic state' of system. For definiteness, a closed, isothermal system at constant volume is considered.

A change of kinetic state within the system may be linked to the presence of a scalar force - the activation affinity. It reflects the difference in activation free energy, $F^{\ddagger}$, of a reaction path at time, t, from its stationary value, some time later (Davis, 1996):

$A^{\ddagger}(i) = -(F^{\ddagger}_{ss}(i) - F^{\ddagger}_{t}(i)) = RT(\ln K^{\ddagger}_{ss}(i) - \ln K^{\ddagger}_{t}(i))$

$$= -RT \ln f^{\ddagger}(i) \qquad\qquad (2)$$

R is the gas constant and T temperature. $K^{\ddagger}$ is an activation state 'equilibrium constant'; $K^{\ddagger}(i) = T(i)/A$. A time-dependent expression relating the $i^{th}$ transition state concentration, $T_t(i)$, to its steady state level, $T_{ss}(i)$, is obtained on solving the rate equations for (1);

$f^{\ddagger}(i) = K^{\ddagger}_{t}(i)/K_{ss}(i) = T_i/T_{ss}$

$$= \underline{\frac{A_0 k_{1,i}}{(k_{2,i} - k_{1,i})}} (e^{-k1,it} - e^{-k2,it}) \Big/ \underline{\frac{A_0 k_1 e^{-k1,it}}{k_{2,i}}}$$

$$= \underline{\frac{k_{2,i}}{(k_{2,1} - k_{1,i})}} (1 - e^{-(k2,i - k1,i)t}) \qquad\qquad (3)$$

$A_0$ is the initial reactant concentration. Steady state requires $k_{2,i} \gg k_{1,i}$, leaving the factor $(1 - e^{-(k2,1 - k1,i)t})$ as a measure, $f^{\ddagger}(i)$, of advancement by the $i^{th}$ reaction path to steady state. When the reaction (1) is catalysed

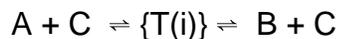

$K_t^{\ddagger}(i)$ is $T_t(i)/(AC)$ and $K_{ss}^{\ddagger}(i)$ equals $T_{ss}(i)/(AC)$, which reduces to $k_1/k_2$. The second order advancement term, $f^{\ddagger}(i)^{(2)}$, is

$$\frac{k_{2,i}}{(k_{2,i} - k_{1,i}(A+C))} \left( 1 - \frac{k_{2,i} - k_{1,i}(A + C)}{k_{2,1} - k_{1,i}(A_0 + C_0)} \ \exp\text{-}[k_{2,i}t - \int_0^\tau dt\ k_{1,i}(A + C)] \right)$$

where $k_{2,i} \gg k_{1,i}(A + C)$.

The steady state constraint, requiring that the first step in (1) be rate limiting, is a feature of the reaction. It implies that the formation of each $T(i)$ involves reactant activation. In the second step of (1), $T(i)$ decomposes rapidly to product, at a rate determined by 'diffusion' along the high energy segment, $\delta$, of the reaction path. The decomposition frequency of the $i^{th}$ transition state is

$$k_{2,i} = k_{-1,i} = \left( \frac{k_B T}{2\pi m_i^\ddagger} \right)^{1/2} \frac{1}{\delta_i}$$

$k_B T$ being the thermal energy per molecule, and $m_i^\ddagger$ is the mass of the $i^{th}$ activated complex. Although $k_{2,i}$ and $k_{-1,i}$ appear equal in (1), reaction rate theory (Eyring and Eyring, 1963) specifies that, once formed, the activated complex proceeds to product. A transmission coefficient less than one being discounted here, and quantum tunnelling through the activation energy barrier is disregarded also. When the reaction (1) proceeds spontaneously in its forward direction, each formation rate coefficient, $k_{1,i}$, exceeds the corresponding reverse coefficient, $k_{-2,i}$. The former being defined here as,

$$k_{1,i} = \frac{(2\pi m_i^\ddagger k_B T)^{1/2} \delta_i Q_i^\ddagger}{h \qquad Q_A} \ e^{-\epsilon_{0,i}/RT}$$

$h$ is Planck's constant. $Q_i^\ddagger$, $Q_A$ are the partition functions of the $i^{th}$ transition state (omitting the translation term; $(2\pi m_i^\ddagger k_B T)^{1/2}/h$) and reactant; $Q_A Q_C$ replaces $Q_A$ for the catalysed reaction path. $\epsilon_{0,i}$ denotes the zero point energy per mole. In addition to the free energy of activation, implicit in the relation for $k_{1,i}$, $k_{-2,i}$ reflects the free energy difference between A and B.

The rate of a reaction, in which all paths are retained during a change of kinetic state, is increased by addition of a path. Path number ordinarily increases with the introduction of a catalyst. Total transition state concentration increases on passing from an uncatalysed, $T_u$, to catalysed, $T_u + T_c$, kinetic state. The change in activation free energy is

$$\Delta F^\ddagger = F^\ddagger_c - F^\ddagger_u = -RT \ln [1 + K^\ddagger_c/K^\ddagger_u] \leq 0 \qquad (4)$$

$K_c^\ddagger = T_c/(AC)$ and $K_u^\ddagger = T_u/A$, where T, A and C are transition state, reactant and catalyst concentration - activity for (4) non-ideal. The inequality applies for all values of the activated 'equilibrium constant' $K^\ddagger$; $K^\ddagger_c \lesseqgtr K^\ddagger_u$. A decrease in overall activation free energy requires only that the number of paths increase; transitions with a large number of paths are conceivable. Conversely, a spontaneous decrease in path number is excluded by (4). This would require that a standard system spontaneously shift to a less probable set of paths. The probability of this transition decreases exponentially with the increase in activation free energy. A spontaneous transition among competitive autocatalytic processes, by contrast, will exhibit a significant reduction in path number. As slow paths are eliminated, the most probable kinetic state then consists solely of the fastest reaction path!

A spontaneous change of kinetic state is characterised by a positive activation affinity. From (2) and (3),

$e^{-A\ddagger(i)/RT} = T_t(i)/T_{ss}(i) \leq 1$ , $A^{\ddagger}(i) \geq 0$ (5)

A positive kinetic force implies a transition state has a probability in the canonical distribution of states that is higher than its pre-steady state probability. With relaxation to the canonical distribution during reaction advancement, transition state probability increases to its steady state level and the kinetic force vanishes. A positive force of path displacement does not represent a potential for chemical work, since system free energy is not altered by changes in path probability between A and B states (1). A change of kinetic state thus involves a novel kind of force: a scalar quantity linked to a displacement, which produces no work or heat.

3. **Force Damping by Pre-Steady State Changes**

It can be demonstrated that changes in path probability in a pre-steady state system optimally damp the activation affinity.

Differentiation of (2) with respect to time reveals that the activation affinity is damped as

$A^{\ddagger}(i)' = -RT\phi_i \leq 0$, $\phi_i \geq 0$ (6)

$\phi_i$ is $(k_{2,i} - k_{1,i})e^{-(k2,i - k1,i)t}/(1 - e^{-(k2,1 - k1,i)t})$ in a first order reaction and

$(k_{2,i} - k_{1,i}(A + C))\, exp\text{-}[k_{2,i}t - \int_0^\tau dt\, k_{1,i}(A + C)]/(1 - exp\text{-}[k_{2,1}t - \int_0^\tau dt\, k_{1,i}(A + C)]$

at second order. As noted earlier (section 2), $k_{2,i} \gg k_{1,i}$ in the former and $k_{2,i} \gg k_{1,i}(A +$

C) in the latter is a precondition for achieving steady state kinetics. In view of (2), the damping rate (6) corresponds to $d(\ln f^{\ddagger}(i))/dt$. Since $f^{\ddagger}(i)$ reduces to $T_t(i)/T_{ss}(i)$, as $T(i)$ approaches a steady state concentration, $T_{ss}(i)$, both the activation affinity (2) and its damping rate (6) diminish and ultimately vanish.

The probability that a reaction will occur by its $i^{th}$ path at time t is

$$p_t(i) = K_t^{\ddagger}(i)/\sum_i K_t^{\ddagger}(i) \tag{7}$$

Its rate of change is

$$p_t(i)' = p_t(i)\phi_i(1 - s_i) \tag{8}$$

In this expression for $p_t(i)'$, $s_i$ is a control variable, whose value determines the direction of change in path probability with time. $s_i$ equals $\phi/\phi_i$, where $\phi$ is $\sum_i \phi_i p_t(i)$. It is demonstrated in Appendix A, by the method of undetermined multipliers, that this rate equation maximises $\phi$ for a path frequency distribution of given variance. It follows from (6) that the mean activation affinity is optimally damped.

## 4. Least Action Path to Steady State

The action specified by mean activation affinity values (2) over time in a pre-steady state standard reaction is

$S = \int_0^\tau dt\ A^\ddagger$

$\quad = -RT \int_0^\tau dt\ \sum_i p_t(i)\ \ln [f^\ddagger(i)]$ \qquad\qquad\qquad (9)

Path frequency changes (8) accompanying the transition to steady state are demonstrated in Appendix B to give a stationary value for the action. It was shown above (section 3) that the activation affinity is minimised at each instant during this transition. It is concluded from this finding that a stationary value for the action (Appendix B) corresponds to the system following a path of least cost in terms of action, while evolving toward steady state.

5. **Analysis of a Pre-Steady State Reaction**

Spectrophotometric measurements of 2,4-dinitrophenol (DNP) formation on phosphorolysis of 2,4-dinitrophenyl phosphate (DNPP), in the presence of alkaline phosphatase (EC 3.1.3.1), using the stop-flow technique (Trentham and Gutfreund, 1968)  illustrate the effect of a change of kinetic state on the kinetic force in a standard reaction.

Figure 1 shows the rate of DNP formation at 25$^o$C and pH 5.9 decreased over time, with an effective first order rate coefficient of nearly 30 s$^{-1}$. A decreased rate of DNP formation accompanied reaction advancement, because dephosphorylation of the enzyme-phosphate complex, E-P,

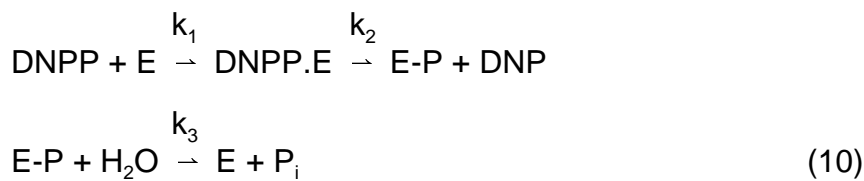

proved to be rate limiting under acidic conditions: $k_3 \ll k_1$ or $k_2$. Consequently, E-P

levels increased at the expense of active enzyme molecules, E. Uncatalysed

hydrolysis of the phosphate bond in DNPP

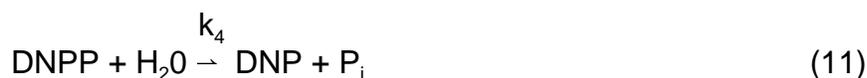

$$DNPP + H_2O \xrightarrow{k_4} DNP + P_i \tag{11}$$

to produce DNP and an inorganic phosphate ion, $P_i$, did not exceed 1 per cent of the catalysed rate.

FIGURE 1

As the transition state in the catalysed reaction, $E\text{-}P.H_2O$, increased toward its

steady state concentration, the activation affinity, $A^{\ddagger}_c$, decreased. From a value of

nearly 14 kJ mol$^{-1}$ at 0.1 ms, the force reduced to only 0.03 kJ mol$^{-1}$ by 150 ms. Its

rate of decrease (6)

$$A^{\ddagger}_c{}' = -RT30e^{-30t}/(1 - e^{-30t}) \tag{12}$$

also diminished with time. The rate was -2440 kJ mol$^{-1}$ s$^{-1}$ at 1 ms. At 10 and 100 ms after the start of the reaction, it reduced to -213 and -4 kJ mol$^{-1}$ s$^{-1}$, respectively. With the uncatalysed reaction (11) remaining at its steady state, the activation affinity for this path, $A^{\ddagger}_u$, was zero throughout the experiment (Fig. 1).

The action based on mean activation affinity (9) values during DNPP

phosphorolysis was evaluated. The probability of catalysed and uncatalysed paths for the pre-steady state interval relied on a steady state probability for the latter of 0.01. An interval of 153 ms was required to reach a catalysed path probability close to 0.99. The action over this interval was 87 kJs mol$^{-1}$.

Results presented earlier (section 2,3) establish that activation affinity damping rates and the action obtained for the DNPP reaction, in its transition to steady state kinetics, are extremal values.

6. **Discussion**

Changes in the path probability distribution of a standard reaction, with multiple paths, could have been investigated at equilibrium. In this state, no thermodynamic force acts on the system and no change occurs in reactant and product concentration. The system is then effectively in a state of maximum symmetry (Callen, 1974, 1985). In particular, it exhibits no time asymmetry of thermodynamic origin. Addition of a catalyst to a chemical system at equilibrium, notwithstanding thermodynamic invariance, elevates the rate of interconversion between reactant and product molecules, to produce an effect similar to that observed on mixing alkaline phosphatase and DNPP (section 5). In the presence of a catalyst the system moves to a more probable kinetic state, as the path probability distribution of a reaction shifts to lower the free energy of activation (section 2).

During this transition, a chemical system at equilibrium changes in a preferred direction with time. Since this time asymmetry arises in the absence of thermodynamic forces, a second source of time asymmetry evidently occurs within the system. A second scalar force was identified: the activation affinity. This finding adds to the number of known sources (Davies, 1974; Penrose, 1979; Schulman, 1996) of time asymmetry.

Both thermodynamic and kinetic time asymmetries derive from statistical displacement of a system. They reflect a probability difference between an initial and final state, and forces of both types are damped as the system moves toward its most probable state or set of reaction paths. At the most probable state consistent with kinetic and thermodynamic constraints, these forces vanish and no further change occurs. The amount of heat contained by a system, at a specified temperature, is then maximised, in accord with the second law of thermodynamics. In addition, kinetic theory requires that the rate of molecular interconversion is also maximised, for a specified set of reaction paths.

The activation affinity displaces a chemical system, without yielding either work or heat. Shifting the path probability distribution between reactant and product state causes no change in system free energy and, therefore, no change in work potential (section 2). The concept of a kinetic force thus generalises the notion of force. It requires inclusion of a scalar quantity responsible for a kind of displacement, which is not coupled to work output. Similar to classical and quantum forces in a conservative dynamical system (Feynman & Hibbs, 1965; Soper, 1976) and the thermodynamic force (Shiner & Sieniutycz, 1994), the kinetic force was demonstrated to move a system along the path of least action (section 3).

Darwinian evolution based on natural selection among competitively replicating RNA molecules (Kramer et al., 1974) was shown to damp the activation affinity for nucleotide condensation (Davis, 1996). This system evolved toward steady state kinetics over an interval of 3 hr. In this time, the effective activation free energy for nucleotide condensation decreased from 86.90 to 86.55 kJ mol$^{-1}$. Consistent with the time dependence of the mean polymer formation rate coefficient (Davis, 1978), the mean activation affinity, varied with the length of the replication interval, L,

$$A^{\ddagger} = 0.33 + 3.27 \times 10^{-5}\ L - 1.35 \times 10^{-8}\ L^2 + 7.15 \times 10^{-13}\ L^3$$

decreasing from 0.33 to 0.0076 kJ mol$^{-1}$ over 190 min. (L = 11400 s). The chemical affinity driving nucleotide condensation during polymerisation was held virtually constant at 32.2 kJ mol$^{-1}$, throughout the transition to steady state kinetics. Darwinian evolution under defined physical conditions thus damped the kinetic force within a chemical system, moving the system toward the most probable (fittest) reaction path. The principle of natural selection is seen to complement Boltzmann's evolution principle, requiring that state-centred forces are damped with time by a statistical mechanism.

**Appendix**

A. **Maximum Rate of Increase in Transition State Concentration**

Changes in reaction path frequency with time in a pre-steady state standard system shall be demonstrated to maximise the rate at which the concentration of each transition state approaches its steady state level.

Recalling (7) and (8), path probability and its rate of change are

$p_t(i) = K^{\ddagger}_t(i)/\Sigma_i\ K^{\ddagger}_t(i)$

$p_t(i)' = p_t(i)\phi_i(1 - s_i), \quad s_i < 1 \ \Rightarrow \ p_t(i)' > 0 \hspace{3cm} \text{(A1)}$
$\hspace{4.7cm} s_i > 1 \ \Rightarrow \ p_t(i)' < 0$

As indicated, $s_i$ controls the direction of change in path probability. $K^{\ddagger}_t$ is the transition state to reactant concentration ratio at time t, and $\phi_i$ is d ln $K^{\ddagger}_t(i)/K^{\ddagger}_{ss}(i)$/dt.

The latter relates changes in the current level of the $i^{th}$ transition state to its steady state value. Should path frequency changes (A1) maximise $\phi$, pre-steady state changes in a standard chemical system would advance transition state levels toward steady state at their maximum rate. This implies maximum damping of the activation affinity (6).

Displacement of the path frequency distribution is subject to two constraints: (i)

the probabilities are normalised, so $\Sigma_i\ \delta p(i) = 0$, and (ii) the variance in $\phi$ is momentarily fixed, $\Sigma_i\ p(i)(\phi_i - \ \bar{\phi})^2 = B$, where B is a constant. Using (A1), constraint

(ii) can be recast as

$\Sigma_i\ \dfrac{\delta p(i)}{\delta t}\ \dfrac{\delta \ln p(i)}{\delta t} = B$

We may then define the function

$$\psi = \Sigma_i \, \phi_i \delta p(i) + \lambda_1 (\Sigma_i \, \frac{\delta p(i)}{\delta t} \, \frac{\delta \ln p(i)}{\delta p(i)} - B \delta t) + \lambda_2 \Sigma_i \, \delta p(i) \qquad (A2)$$

It is stationary with respect to variations in probability, $\delta p(i)$, when

$$\frac{\partial \psi}{\partial \delta p(i)} = \phi_i + 2\lambda_1 \frac{\delta p(i)}{p(i)\delta t} + \lambda_2 = 0 \qquad (A3)$$

The Lagrange multipliers are found to be $\lambda_1 = -\frac{1}{2}$ and $\lambda_2 = -\bar{\phi}$. Substitution into (A3) reveals

$$(\delta p(i)/\delta t)_{max} = p(i)(\phi_i - \bar{\phi}) \qquad (A4)$$

The probability variation is a maximum, as differentiation of (A2) again with respect to $\delta p(i)$ reveals the second derivative is negative:

$$\partial^2 \psi / \partial \delta p(i)^2 = - l/p(i)\delta t \leq 0, \; 0 \leq p(i) \leq 1 \text{ and } \delta t \geq 0.$$

B. **Transition Path of Minimum Action**

An action (cost function) for the pre-steady state transition by a standard reaction with multiple paths (1) is given by

$$S = - RT \int_0^\tau dt \, \Sigma_i \, p_t(i) \ln [f^{\ddagger}(i)] \qquad (B1)$$

the integral of the mean activation affinity (2) over the transition interval. During this interval, the path probability distribution $(p_t(1), p_t(2),...)$ approaches its steady state (Boltzmann) distribution. Using Pontryagin's minimum principle (PMP) from control theory (Pontryagin et al., 1962), changes in path probability specified by the state equation (A1) will be shown to yield a stationary value for the action (B1).

The cost variable adjoined to the state equation (A1) is

$$p_t(0)' = - RT \sum_i p_t(i) \ln [f^{\ddagger}(i)] \qquad (B2)$$

The resulting Hamiltonian from (A1) and (B1) is

$$H = - RT \lambda_0 \sum_i p_t(i) \ln [f^{\ddagger}(i)] + \sum_i \lambda_i p_t(i)\phi_i(1 - s_i) \qquad (B3)$$

The Hamiltonian relations then yield

$$\lambda_0' = - \partial H/\partial p_t(0) = 0 \;\; \Rightarrow \;\; \lambda_0 = 1 \;\; (PMP)$$

$$\lambda_i' = - \partial H/\partial p_t(i) = RT \ln [f^{\ddagger}(i)] - \lambda_i\phi_i(1 - s_i) \qquad (B4)$$

$\lambda_0$ is set equal to 1 on applying PMP, after Hamilton's relations show it is a constant. Solving the ODE in (B4) for $\lambda_i$ time dependent yields

$$\lambda_i(t) = \exp [-\int_0^t \phi_i(1 - s_i)dt'] \int_0^t RT \ln [f^{\ddagger}(i)] \exp [\int_0^t \phi_i(1 - s_i)dt'] \, dt$$

Substitution for $\lambda_i$ in (B3), for all i, gives a time dependent Hamiltonian. Since it is not constant for the pre-steady state transition path of the system, it is discounted as a non-optimal path (Pontryagin et al., 1962).

For $\lambda_i$ independent of time ($\lambda_i' = 0$), (B4) gives (Davis, 1996)

$$\lambda_i = \frac{RT \ln [f^\ddagger(i)]}{\phi_i(1 - s_i)} \qquad (B5)$$

Both $RT \ln [f^\ddagger(i)/\phi_i(1 - s_i)$ and its time variation $(d(RT \ln [f^\ddagger(i)])/dt/d(\phi_i(1 - s_i)/dt)$ are equal to some constant, c. Recalling the relation between $\phi_i(1 - s_i)$ and the rate of change in path probability, $p(i)'$, in (A1) and the definition of the kinetic force (2), it is apparent from (B5) that

$$c \, d \ln p(i)/dt = - A^\ddagger_i \qquad (B6)$$

With integration

$$p_t(i) = p_0(i) \exp [\textstyle\int_0^t A^\ddagger_i(t)/RT \, dt] \qquad (B7)$$

Here, the probability of a reaction path at some time, t, during the transition to steady state depends on its initial probability, $p_0(i)$, and the applied kinetic force, $A^\ddagger_i(t)$, over the transition interval (0,t). The constant c has been equated with -RT, following inspection of (B5) and on noting that $f^\ddagger(i)$ measures advancement of the i[th] reaction path toward steady state or, equivalently, changes in probability of the i[th] reaction path. Substitution of the time invariant costate variable, $\lambda_i$, in the Hamiltonian (B3) reveals

$$H/RT = - \sum_i p_t(i) \ln [f^\ddagger(i)] + p(1) \ln [f^\ddagger(1)] + p(2) \ln [f^\ddagger(2)] + \ldots \quad (B6)$$

$$H = 0$$

The control variable, $s_i$, specified by (B5), therefore, yields a stationary Hamiltonian. Control theory indicates that a zero-valued Hamiltonian, over an interval with a freely chosen final time, $\tau \geq 0$, corresponds to a stationary value of the cost function. As (B1) is the cost function for the pre-steady state transition, the path traced yields a stationary action. It was previously established (Appendix A) that the activation affinity, on which this action (B1) is based, is minimised at each instant throughout the transition interval. Consequently, a pre-steady state system can be concluded to follow a path of least action.

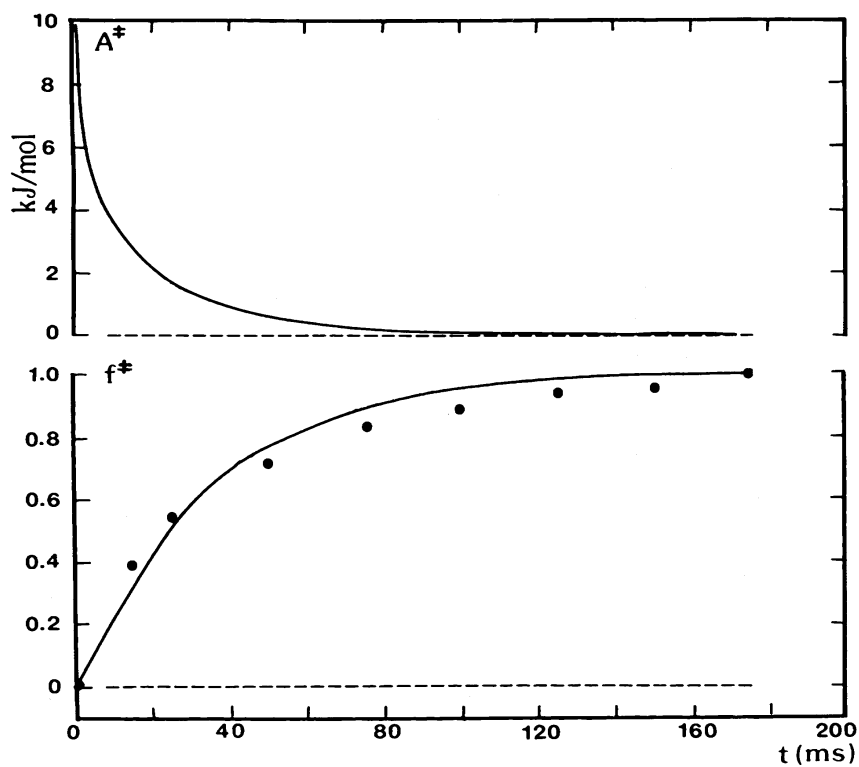

Figure 1. Damping of the activation affinity, $A^{\ddagger}$, by the shift to steady state kinetics

during phosphorolysis of DNPP. Reaction advancement was determined by

spectrophotometric measurements (Trentham and Gutfreund, 1968) on DNP

liberation over an interval of 175 ms, after initiation of DNPP hydrolysis by alkaline phosphatase at $25^{\circ}$C and pH 5.9 with the stop-flow technique. Continuous line,

enzyme catalysed. Broken line, uncatalysed. Points, transition state concentration

as a fraction, $f^{\ddagger}$, of its steady state concentration. These values were calculated for

indicated times, t, on the spectrophotometric record of DNP release. The $f^{\ddagger}$ values

obtained correlate with the time course of a first order process (continuous line)

having a rate coefficient of $30s^{-1}$ ($r^2 = 0.976$).